\documentclass[conference]{IEEEtran}
\IEEEoverridecommandlockouts
\usepackage{cite}
\usepackage{amsmath,amssymb,amsfonts}
\usepackage{algorithmic}
\usepackage{graphicx}
\usepackage{textcomp}
\usepackage{xcolor}
\def\BibTeX{{\rm B\kern-.05em{\sc i\kern-.025em b}\kern-.08em
    T\kern-.1667em\lower.7ex\hbox{E}\kern-.125emX}}
\usepackage{enumitem}
\usepackage{cleveref}
\usepackage{amssymb}
\usepackage{multirow}
\usepackage{array}

\newcommand{\CE}{Continuous Experimentation}
\newcommand{\cps}{cyber-physical systems}

\newboolean{showcomments}
\setboolean{showcomments}{true} 
\ifthenelse{\boolean{showcomments}}
  {
		\newcommand{\nbb}[2]{
		\fcolorbox{black}{yellow}{\bfseries\sffamily\scriptsize#1}
		{\sf$\blacktriangleright$\textcolor{blue}{\textit{#2}}$\blacktriangleleft$}
		}

  }
  {
		\newcommand{\nbb}[2]{}
  }

\usepackage{tikz}

\newcommand\copyrighttext{%
  \footnotesize \textcopyright 2019 45th Euromicro Conference on Software Engineering and Advanced Applications (SEAA)}
\newcommand\copyrightnotice{%
\begin{tikzpicture}[remember picture,overlay]
\node[anchor=south,yshift=30pt] at (current page.south) {\fbox{\parbox{\dimexpr\textwidth-\fboxsep-\fboxrule\relax}{\copyrighttext}}};
\end{tikzpicture}%
}

\begin{document}

\title{The Automotive Take on \CE: A Multiple Case Study
}

\author{\IEEEauthorblockN{Federico Giaimo}
\IEEEauthorblockA{
\textit{Chalmers $|$ University of Gothenburg}\\
Gothenburg, Sweden \\
giaimo@chalmers.se}
\and
\IEEEauthorblockN{Hugo Andrade}
\IEEEauthorblockA{
\textit{Chalmers $|$ University of Gothenburg}\\
Gothenburg, Sweden \\
sica@chalmers.se}
\and
\IEEEauthorblockN{Christian Berger}
\IEEEauthorblockA{
\textit{University of Gothenburg}\\
Gothenburg, Sweden \\
christian.berger@gu.se}
}

\maketitle
\copyrightnotice

\begin{abstract}
Recently, an increasingly growing number of companies is focusing on achieving self-driving systems towards SAE level 3 and higher. 
Such systems will have much more complex capabilities than today's advanced driver assistance systems (ADAS) like adaptive cruise control and lane-keeping assistance. 
For complex software systems in the Web-application domain, the logical successor for Continuous Integration and Deployment (CI/CD) is known as Continuous Experimentation (CE), where product owners jointly with engineers systematically run A/B experiments on possible new features to get quantifiable data about a feature's adoption from the users.  
While this methodology is increasingly adopted in software-intensive companies, our study is set out to explore advantages and challenges when applying CE during the development and roll-out of functionalities required for self-driving vehicles. 
This paper reports about the design and results from a multiple case study that was conducted at four companies including two automotive OEMs with a long history of developing vehicles, a Tier-1 supplier, and a start-up company within the area of automated driving systems. 
Unanimously, all expect higher quality and fast roll-out cycles to the fleet; as major challenges, however, safety concerns next to organizational structures are mentioned. 

\end{abstract}

\begin{IEEEkeywords}
continuous experimentation, automotive, cyber-physical systems, autonomous driving
\end{IEEEkeywords}

\section{Introduction}

In recent years the automotive industry has witnessed a rapid revolution in the way its products are created and what functionality can be offered to their customers. 
Software is nowadays one of the main components of any vehicle~\cite{URL_FUSE_Hiller}, enabling more and more functions necessary to achieve automated driving capabilities in the foreseeable future. 

In order to be competitive in this scenario it is important to roll out changes and new software as quickly as possible. Such a routine would require less effort in a context where safety was not as fundamental, but due to the risks involved in the automotive field, safety regulations rightfully impose important restrictions on what hardware and software are allowed on public roads. 
An effect of this is the need for lengthy testing and verification processes. 
ISO 26262
~\cite{URL_iso26262} and ISO 21448
~\cite{URL_iso21448} are two of such standards, involved in the definition of the functional safety requirements of vehicles.

A possible way to accelerate the release of new software is to make use of the data that are already available and that are generated during a vehicle's lifetime. 
An engineering practice with this aim is \CE, which enables the product owner to deploy additional experimental software alongside its official version to obtain data and practically verify its performance with respect to specified evaluation criteria. 
The experimental software can be run in closed- or open-loop mode, depending on whether it is the experimental or the official software to have control over the system. 

While \CE\ is widespread in settings like web-based software-intensive systems~\cite{KDF+13}\cite{TAOM10}, it is still not largely applied in the context of embedded and cyber-physical systems such as the automotive field~\cite{RR18}. 
A reason for this is the possibility to introduce safety risks to road users, not only because of the experimental nature of the software under test, but also because of the few available computational resources. 
The physical limits of the available resources introduce in their own right additional challenges to the adoption of \CE\ and will require ad-hoc solutions, as highlighted in a previous study~\cite{GBK17}. 

This paper aims at reporting the feedback, impressions, 
and expectations from industrial representatives that participated in a series of workshops focusing on what \CE\ can offer to them in terms of advantages and expected challenges with respect to their role in their respective companies. In this sense, the goal of this work can be expressed through the main questioning: \textit{How desirable is the Continuous Experimentation practice to automotive practitioners and what are the obstacles that they perceive are preventing its adoption in the industrial field?}

This work contributes to the current body of knowledge with a depiction of the present \textit{understanding} (the term is used as opposed to \textit{state-of-practice}) of \CE\ in the automotive field, grounded in empirical data. 
The principal observation emerging from the study is that \CE\ is perceived as a positive practice capable of bringing to this field the same advantages that it has brought to web world, but its realization is opposed by both significant technical challenges and a conservative organizational and legal framework, making the automotive industry fall behind with what concerns the adoption of this practice. 



Although some of the causes for this delay can be intuitively inferred, this study has been devised to clearly understand the reasons by involving the practitioners and investigating both the positive expectation and the challenges that this practice can pose for their professional role.

\section{Research Method} \label{sec:RM}


In order to engage and discuss with industrial representatives in relevant roles, such as technical leaders and supervisors, a series of workshops was organized by the authors, one for each of the companies. 
The theme of the workshops was the introduction of the concept of \CE\ to automotive representatives to provide a common vocabulary for the workshop in order to obtain their feedback on two questions, derived from the main questioning:
\begin{enumerate}[leftmargin=*,align=left,label=\textit{Q{\arabic*}:}]
\item \textit{What would be the added values for your role in the context of self-driving vehicles if \CE\ is successfully in place?}
\item \textit{What would be the additional challenges for your role in the context of self-driving vehicles if \CE\ would be in place?}
\end{enumerate}
The resulting answers and discussions helped to clarify the state-of-practice of \CE\ and the prospects and obstacles that are still perceived in the way for its adoption in the companies that participated in the workshops. 

\subsection{Involved companies} 
Four companies were involved in this study. 
They were chosen due to their known efforts towards autonomous driving. 
Additionally, they constitute an interesting set due to their diversity, as they comprise two automotive OEMs (Original Equipment Manufacturer) in this article named Companies A and B, a Tier-1 supplier named Company C, and an autonomous driving start-up company named Company D.


%

The OEMs have a years-long history of developing consumer vehicles and recently increased activities in their research around highly automated driving solutions. 
Complementary to the OEMs, an established tier-1 supplier was also contacted. 
This company has a years-long experience in supplying safety systems for automotive OEMs and has increased its activities towards components and solutions for active safety systems and automated driving solutions. 
In contrast to these companies, one workshop was also run at a start-up company working on an autonomous electric vehicle, which does not have to obey to a large legacy code-base grown over the years but could nearly start from scratch adopting the most suitable processes.



The participants were chosen among diverse technical roles in the companies. 
Among the overall set of attendees there were developers, software architects, team leaders, and a manager.
The variety of roles was reputed a positive factor due to the fact that the workshop questions related to the participants' role in their respective companies.
An increased diversity of represented areas was hence considered an element providing additional perspectives in the answers and discussion. 
The participants' roles were: from Company A, 3 software developers, 1 team leader and 1 manager; from Company B, 2 software developers and 2 team leaders; from Company C, 1 software developer and 2 team leaders; lastly, from Company D, 1 software developer and 1 team leader.



\subsection{Format of the workshops}
Each workshop lasted in total between 1.5 and 2 hours, depending on the number of participants. During the workshops, one of the authors would lead it through its different phases, while the other authors would assist and take notes. 
The format was organized in four phases as follows:

\textbf{Phase I:} After each participant would have presented himself and his role to the group, an initial presentation was shown. 
The goal of the presentation was to establish a common understanding and vocabulary of the Continuous practices, namely Continuous Integration, Continuous Delivery/Deployment, and \CE. This phase would take around 20 minutes; 

\textbf{Phase II:} At the end of the presentation, the participants were asked the two aforementioned questions. 
They were given time to individually devise their answers, writing each idea on a note. This phase would take around 30 minutes; 

\textbf{Phase III:} The participants took turns to go through their notes in order to explain to the group their meaning and the reasoning behind it. Each note would then be placed near others on the same theme on a whiteboard, thus creating thematic clusters. This phase would take around 40 minutes; 

\textbf{Phase IV:} An infrastructure model for \CE\ devised for companies with web-based products~\cite{FGMM17} was presented to the participants. They were asked to jointly discuss the model with the aim of identifying critical points and necessary changes if it had to be applied to the automotive industry. This phase would take around 15 minutes.


The described format with open questions, which are locked on a structured topic, categorizes the workshop series as semi-structured case study~\cite{RH09}. 
This approach was chosen because it fits well the exploratory and explanatory goal of this work by promoting the participants to provide original feedback, ideally completely unbiased by the authors. 

\subsection{Data collection and handling}

The notes and comments of the industrial representatives were transcribed and used to identify common themes among advantages and challenges. 
To improve the quality of the data and increase the validity of our results the ``observer triangulation'' was implemented whenever possible, meaning that there was more than one observer collecting data and feedback~\cite{RH09} in all workshops with the exception of the one organized with the start-up company representatives. 

At the end of the data collection the transcriptions and the raw data,~i.e., the notes produced by the participants, were re-examined and discussed to ensure a common understanding among the authors and an accurate representation in the present article. 

\section{Results and Discussion}  \label{sec:R}

The findings are a result of a bottom-up approach in which topics of interest from the discussions in the workshops were identified. 
The topics take into account the data obtained from the four companies and were combined into a comprehensive list. 
They are organized and described according to the two research questions that guided this study: the Added Values of \CE\ answer Q1 and are summarized in Table~I, while the Challenges \CE\ will pose assess Q2 and are reported in Table~II.

\begin{table*}[ht]
\begin{tabular}{m{1.5cm} p{2.5cm} p{11cm} c}
\textbf{Category} & \textbf{Value} & \textbf{Description} & \textbf{Companies} \\ \hline
\multirow{3}{*}[-2.28em]{Safety} & \multirow{1}{*}[-0.5em]{Monitoring} & Allows for constant notifications about software issues, therefore leading to quicker fixes. Developers can also obtain a better understanding of the user interaction and system behavior. & \multirow{1}{*}[-0.5em]{B, C, D}\\ \cline{2-4}
& \multirow{1}{*}[-1.1em]{Mechanical integrity} & Constant monitoring result in a slower wear and tear of mechanical components by interpreting situational/behavioral states of the system. Once identified, wear-prone situations could be avoided. & \multirow{1}{*}[-1.1em]{C, D} \\\cline{2-4}
& \multirow{1}{*}[-0.5em]{Easier testing} & Field testing on the fly makes it easier to detect bugs, and with the constant feedback it would be easier to find relevant test cases for the system. & \multirow{1}{*}[-0.5em]{A, B, C}\\ \hline
\multirow{2}{*}[-2.85em]{Speed} & \multirow{1}{*}[-1.15em]{Faster data collection} & Relevant data can be collected on demand, rather than from controlled tests, allowing for fast analysis of system behavior. OEMs can benefit from the real-world system usage due to the OTA connection. & \multirow{1}{*}[-1.15em]{A}\\ \cline{2-4}
& \multirow{1}{*}[-1.7em]{Faster time-to-market} & Software can be updated regularly, without manual delivery of new versions. Instead of typical acceptance testing with a reduced number of users, the acceptance can be measured from real-world scenarios as fast as the data can be transmitted to the headquarters. Further, developers can avoid ``big bang'' integration by incrementally adding features. & \multirow{1}{*}[-1.7em]{A, B, C, D}\\ \hline
\multirow{1}{*}[-0.55em]{Quality} & \multirow{1}{*}[-0.5em]{Customer satisfaction} & Functionalities are reassessed using data from regular usage. The customers' preferences are captured and implemented into the system through updates, improving customer satisfaction. & \multirow{1}{*}[-0.5em]{B, C, D}\\ \hline
\multirow{1}{*}[-1.15em]{Sustainability} & \multirow{1}{*}[-1.1em]{Energy efficiency} & Unused functionalities can be disabled to reduce energy consumption. The data resulting from a constant monitoring of the hardware’s energy consumption can also be used to improve energy efficiency. & \multirow{1}{*}[-1.1em]{D}\\ \hline
\multirow{3}{*}[-1.65em]{Opportunities} & Real-world data usage & Learning from data enables research and improvements of both the process and the product. Further, the collected data can be analyzed and/or sold as services. & \multirow{1}{*}[-0.5em]{A, B, C}\\ \cline{2-4}
& \multirow{1}{*}[-0.5em]{Incremental delivery} & Large and complex functions can be delivered step-by-step. Certain functions can be implemented and updated at a later time. & \multirow{1}{*}[-0.5em]{C}\\ \cline{2-4}
& \multirow{1}{*}[-0.5em]{Fleet view} & Companies may have the opportunity to obtain a comprehensive view of the behavior of their products based on the collected data from the fleet. & \multirow{1}{*}[-0.5em]{A}\\ \hline \\
\end{tabular}
\caption{Added values of \CE\ (Q1) reported in the workshops}
\label{tab:AV}
\end{table*}

\begin{table*}[ht]
\begin{tabular}{m{1.5cm} p{2.5cm} p{11cm} c}
\textbf{Category} & \textbf{Challenge} & \textbf{Description} & \textbf{Companies} \\ \hline
\multirow{3}{*}[-3.4em]{Safety} & \multirow{1}{*}[-1.7em]{Impact measurements} & Measurements must occur before the deployment phase, i.e., the real impact of changes are not entirely under control. Testing is a challenge, e.g., experiments that affect the control of the vehicle. Further, changes in the user experience (e.g., user preferences) may not be appreciated by users. & \multirow{1}{*}[-1.7em]{B, C, D}\\ \cline{2-4}
& \multirow{1}{*}[-1.15em]{Fallback plan} & In case of failure, a fallback plan must always be ready. With multiple versions of the software deployed, this solution demands a robust versioning system that allows safe rollback in case of emergencies. & \multirow{1}{*}[-1.15em]{D}\\\cline{2-4}
& \multirow{1}{*}[-0.5em]{Regulations} & Complying with strict governmental regulations (e.g., in the automotive domain) can be a challenge. & \multirow{1}{*}[-0.5em]{A, B, C, D}\\ \hline
\multirow{2}{*}[-1.2em]{Security} & \multirow{1}{*}[-1.25em]{\shortstack[l]{Data protection \&\\ privacy}} & Major concern since information will move to and from the vehicle. The integrity of the transmission must be preserved through security mechanisms that reduce the risk of interception, impersonation, or tampering. Further, customers might not want to be monitored or participate in experiments. & \multirow{1}{*}[-1.7em]{A, B, C, D} \\ \hline
\multirow{2}{*}[-1.3em]{DevOps} & \multirow{1}{*}[-0.5em]{Versioning} & Developers must acknowledge/monitor versions that are deployed. Different configurations of the same software may be deployed and running on different vehicles. & \multirow{1}{*}[-0.5em]{A, C}\\\cline{2-4}
& \multirow{1}{*}[-0.5em]{Data management} & Collecting, structuring, and analyzing data becomes an integral part of the development process. Only relevant data should be managed rather than excessively large amounts. & \multirow{1}{*}[-0.5em]{A, B} \\ \hline
\multirow{4}{*}[-2.3em]{\shortstack[l]{Quality Assur-\\ance}} & \multirow{1}{*}[-0.5em]{Performance} & Running various instances of the software can be very demanding to the automotive hardware, which is typically resource-constrained. & \multirow{1}{*}[-0.5em]{C, D}\\ \cline{2-4}
& \multirow{1}{*}[-0.5em]{Validation} & Validating software against standards such as the ISO 26262 can be challenging in such highly dynamic environments. & \multirow{1}{*}[-0.5em]{A, B, C, D}\\ \cline{2-4}
& \multirow{1}{*}[-0.5em]{Remote execution} & The risk of unwanted or unknown behavior of the system is increased. Moreover, updates could be at risk of not occurring due to poor, faulty, or non-existing network connections. & \multirow{1}{*}[-0.5em]{C}\\ \cline{2-4}
& \multirow{1}{*}[-0.5em]{Testing} & Since most of the testing in the automotive industry is done manually, this stage represents very high costs. Further, developers may question ``\textit{what is enough testing?}''. & \multirow{1}{*}[-0.5em]{A}\\ \hline
\multicolumn{1}{l}{\multirow{2}{*}[-1.4em]{Costs}} & \multirow{1}{*}[-0.5em]{Hardware} & Additional hardware represents an increase in the cost of the product. Such cost needs to be properly justified by the returns. & \multirow{1}{*}[-0.5em]{A}\\ \cline{2-4}
& \multirow{1}{*}[-0.5em]{Data handling} & Managing large amounts of data reflects on elevated costs e.g., costs for storage and transmission of the data collected by the systems in the fleet. & \multirow{1}{*}[-0.5em]{A}\\ \hline
\multicolumn{1}{l}{\multirow{2}{*}[-0.6em]{Hardware}} & Resource constraints & A highly resource-constrained computational environment limits the options for experimentation. & A, B, C\\ \cline{2-4}
& \multirow{1}{*}[-0.5em]{Heterogeneity} & Systems with different hardware specifications pose a challenge in ensuring that new software versions are supported by the available hardware platforms with their different setups. & \multirow{1}{*}[-0.5em]{B, C}\\ \hline \\
\end{tabular}
\caption{Challenges of \CE\ (Q2) reported in the workshops}
\label{tab:Ch}
\end{table*}

An analysis of the main trends and notes of interest that emerge from the collected data, organized by the research questions from which they originate, follows now. 

\subsection{Added values (Q1)} 
As shown in Table~I, the analysis of the results from the workshop yields that the most mentioned added values, i.e., values that appeared in at least three workshops out of four, were \textit{Easier testing}, \textit{Faster time-to-market}, \textit{Customer satisfaction}, \textit{Real-world data usage}, and \textit{Monitoring}. 
These values hint at the companies' desire for improvements of the testing processes by means of employing data coming from the application field and from the platforms themselves, with the aim to release new software in a faster way and raise its quality while aligning it more to the desires of the customers.
In particular, \textit{Faster time-to-market} was mentioned by all companies, showing that this is the most desired result expected by the adoption of experimentation practices. 

A point of interest emerged from one company, which mentioned the advantages of gaining a \textit{Fleet view} over their vehicles. 
This highlights how even projects relatively small in numbers can introduce a more holistic vision in companies when the products allow them to monitor and influence their behavior in relation to the environment. 



Additionally, it is interesting to notice how Company D 
mentioned as advantages the monitoring of \textit{Mechanical integrity} of the physical elements and the improvement of \textit{Energy efficiency} thanks to optimized guidance software. 
This could hint at a different view of the role of the software as part of the final vehicle: while companies with legacy products may see the software as one of the many components comprising the vehicle and a process to go through to achieve the final product, the autonomous electric vehicle company alludes at the software as the main actor in the product, capable of influencing deeply 
its well-being and lifetime maintenance.

\subsection{Challenges (Q2)} 
The results highlight how the most mentioned challenges, i.e., challenges that were mentioned in at least three workshops out of four, were \textit{Impact measurements}, \textit{Regulations}, \textit{Access to data}, \textit{Validation}, and \textit{Resource constraints}, as shown in Table~II. 

This aligns with the expected obstacles for a practice that involves experimental software in vehicles, which are 1) the possibly negative effects that such software would have on the safety requirements of the vehicle, risking its compliance to existing legal frameworks, as well as 2) the protection of the connected vehicles from malicious third parties that could exploit data or software transmissions to inject unsafe code into the system and 3) the technical difficulty of running a \CE\ process on a platform such as a vehicle, which is equipped with limited computational resources. 

Another interesting note can be extrapolated from the input provided by one company which mentioned the costs of \textit{Data handling}. 
This relates to their aforementioned point about \textit{Fleet view}, as in a pilot project they had a team whose job was to periodically physically access specifically equipped platforms to collect their recorded data. 
The need for such a manual work highlights how important the data retrieval process is, and how expensive it could become when remote or automated procedures are not in place. 

An additional interesting point is the one raised only by Company D, which is the presence of a \textit{Fallback plan}. 
That is necessary for their platform as human accessible maintenance is very limited. 
In this setting there are no vehicle occupants that could take control in case of software failure.



\subsection{Related works}
A number of experience reports and studies have been conducted on \CE\ in the past years. 
Recent mapping studies show how the almost totality of these works is set in the domain of web-based systems, which is the field where the practice originated~\cite{RR18}~\cite{AF18}. 

Very few studies on \CE\ have been performed in the field of embedded systems, cyber-physical systems, or the automotive industry. 

A study linking post-deployment data (although experiments are not explicitly mentioned) and the cyber-physical and automotive field is the one by Olsson and Bosch~\cite{OB13}. 
In their study they interview representatives from three companies, one of which is an automotive manufacturer. 
They show that while post-deployment data collection mechanisms are in place, the collected data is only partially used. 
Moreover, they report that the purpose of this feedback is normally just troubleshooting and not supporting a product improvement process. 

A recent study is the on by Mattos et al.~\cite{MBO18}, which investigates the challenges that embedded systems companies could face when applying \CE. The challenges are identified by analyzing the literature on \CE, which focuses for the most part on web-based systems, and comparing the ones found therein with industrial representatives in order to verify whether those challenges apply to their companies as well. 
A number of their identified challenges overlap with the ones found in the present study, e.g., the potential difficulty and costs to be faced in order to perform effective tests on an experiment-capable platform, and the safety concerns that running experimental software on a safety-critical system would raise. 
However there are also several challenges which are not present here, due to the way their set of challenges was defined, e.g., the lack of experimentation tools integrating with their existing engineering tools, and the need for skilled personnel to tune the experimental software to get meaningful results. 
The present work differentiates from the aforementioned one in both the industrial scope and the format of the case studies, as in the present article the list of challenges are produced by the industrial representatives themselves, and the industry scope is restricted to the automotive field. 
The difference in the way the challenges are produced is the likely reason for the dissimilarities in the challenge sets of the two studies. 
Despite the differences, there is agreement on key issues that make the adoption of \CE\ by the \cps\ industry a complex process.



\subsection{Threats to Validity}
A first possible threat to the validity of this work is rooted in the format of the workshops, as it cannot be ruled out that the presentation run during the initial phase of the workshop biased the participants into focusing mainly on a subset of the possible themes. 
To prevent this, the authors avoided whenever possible to give examples or descriptions of this practice use that could suggest practical applications in the participants' field in order to avoid biases. 
Another threat is the fact that one of the workshops was conducted with only one observer, thus threatening the observer triangulation. 
The counter-measure in this case was to document the workshop as carefully as possible and to later review the notes thoroughly with the other colleagues involved in the study. 
One additional threat to the validity of the conclusions is the fact that was not possible to perform data triangulation,~i.e., to run the same workshops again after a certain period of time to confirm the findings, due to the limited availability of the workshops' participants. 
Finally, another possible threat is the low number of participants, although in the case of Company D this can be explained with the fact that it is a start-up company with a low number of employees suitable for the workshop format. 
However, the limited number of people and companies involved means that the results may not be generalizable to other automotive companies or industrial contexts. Additional workshops will be conducted to verify whether these findings hold true for other companies as well. 

\section{Conclusions and Future Work} \label{sec:CFW}


With this work the authors report the results of a series of workshops executed with automotive representatives. 
The workshops introduced the concept of \CE\ to the participants and asked them to describe what can this practice offer to their role in industry in terms of advantages and improvements to their work but also current or expected challenges.

The results of the workshop confirm that many advantages of \CE, which are well-known in different fields of application, could also be transferred to the automotive field. 
Such advantages include, among others, reducing the time-to-market of new functionality and improving the effectiveness of the software tests using real-world data. 
However, several challenges are still present, such as the possibility to introduce safety risks by adding additional computational load to the system 
and the difficulty to adapt the paradigm shift of this new practice to the existing legal and organizational framework. 

The results describe for Companies A, B, and C a generally interested but conservative attitude, which can be expected of established companies in a highly regulated field facing a novel approach to their work. 
It can be expected however that in time different companies, possibly as Company D is doing, will align their way of working to take systematically advantage of the real-world data that could be available thanks to \CE, in an industry-wide productive trend similar to the one recently witnessed in web-based systems. 

As future steps, additional workshops are planned with other companies in order to further validate and expand the study, providing additional insights into what the industry considers viable, useful, or challenging in the \CE\ practice. 

\section*{Acknowledgment}
This work was supported by the COPPLAR Project -- CampusShuttle cooperative perception and planning platform~\cite{URL_copplar}, funded by Vinnova FFI, Diarienr: 2015-04849.

The authors wish to thank Hang Yin for his availability and help during and after the workshops, and all the industrial representatives for their time and feedback. 

\bibliographystyle{IEEEtran}  

\end{document}